\documentclass[12pt,dvipdfmx]{article}

\renewcommand{\thefootnote}{\fnsymbol{footnote}}

\usepackage{amsbsy,amssymb,latexsym,amsfonts,amsmath}
\usepackage{mathrsfs}
\usepackage[]{graphicx}
\usepackage{bm}
\usepackage{here}
\usepackage{color}

\numberwithin{equation}{section}
\newcommand{\bel}[1]{\begin{equation}\label{#1}}                     
\newcommand{\bal}[1]{\begin{eqnarray}\label{#1}}                     
\newcommand{\be}{\begin{equation}}
\newcommand{\ee}{\end{equation}}

\begin{document}
	%
	%
\begin{titlepage}
		\begin{flushright}
			\normalsize
			~~~~
			NITEP 81\\
			OCU-PHYS 524\\
			February, 2021\\
		\end{flushright}
		
		\vspace{15pt}
		
		\begin{center}
			{\LARGE  Static force potential of}\\
			\vspace{5pt}
			{\LARGE  non-abelian gauge theory at a finite box}\\
			\vspace{5pt}
			{\LARGE  in Coulomb gauge}
		\end{center}
		
		\vspace{23pt}
		
\begin{center}
{Tomohiro Furukawa$^{b}$\footnote{e-mail: furukawa@sci.osaka-cu.ac.jp},
Keiichi Ishibashi$^{b}$\footnote{e-mail: ishibashi.mathphys@gmail.com},}\\
{H. Itoyama$^{a,b,c}$\footnote{e-mail: itoyama@sci.osaka-cu.ac.jp}
and 
Satoshi Kambayashi$^{b}$\footnote{e-mail: kambayashi@zy.osaka-cu.ac.jp}}\\
			
			%
			\vspace{10pt}
			%
			
			$^a$\it Nambu Yoichiro Institute of Theoretical and Experimental Physics (NITEP),\\
			Osaka City University\\
			\vspace{5pt}
			
			$^b$\it Department of Mathematics and Physics, Graduate School of Science,\\
			Osaka City University\\
			\vspace{5pt}
			
			$^c$\it Osaka City University Advanced Mathematical Institute (OCAMI)
			
			\vspace{5pt}
			
			3-3-138, Sugimoto, Sumiyoshi-ku, Osaka, 558-8585, Japan \\

		\end{center}
		%
		\vspace{10pt}
		\begin{center}
			Abstract\\
		\end{center}
		 
		 Force potential exerting between two classical static sources of pure non-abelian gauge theory in the Coulomb gauge is reconsidered at a periodic/twisted box of size $L^3$.  Its perturbative behavior is 
		 examined by the short-distance expansion as well as by the derivative expansion. The latter expansion to one-loop order confirms the well-known change in the effective coupling constant at the Coulomb part as well as the Uehling potential while the former is given by the convolution of two Coulomb Green functions being non-singular at $\bm{x}=\bm{y}$. The effect of the twist comes in through its Green function of the sector.
		 
		
		\vfill
		
\end{titlepage}
	
\renewcommand{\thefootnote}{\arabic{footnote}}
\setcounter{footnote}{0}

\section{Introduction}

The force potential between two static classical sources is a classic object in quantum field theory since Yukawa. In theory where the gauge principle is operating,  the computation of this quantity at the Coulomb gauge is a most straightforward one as the Coulomb potential is present in the interaction Hamiltonian as its instantaneous part\footnote{There is a vast amount of literature dealing with Coulomb gauge non-abelian gauge theory. We give here some of the references \cite{1980CL,1997CZ,2003Zwanziger,2004GOZ,2004RF,2005AT,2006Niegawa,2007WR,2011AT,2012BLetal,2018RBCEH}}. In the covariant gauge, the Coulomb part and the longitudinal part come together in computation and one often derives the potential by comparing it with the nonrelativistic potential in quantum mechanics at the level of amplitudes. 

Non-abelian gauge theory formulated in a finite box has been exploited in several directions both for the periodic 
boundary condition (see, for example, \cite{1983L}) and for the twisted boundary conditions
 \cite{1979tHooft,1981tHooft,1986PvB,1987HPvBZ,1988KPvB,1991PvB,1980tHooftLecture}, combining them 
with several approximations\footnote{For a review, see, for example, \cite{2000PvB}. Also, for Witten index in supersymmetric gauge theories and its computation at finite volume, see \cite{1982Wittenindex,1986S}.}.

The goal of this paper is rather modest: we will reexamine the force potential of the non-abelian gauge theory in the Coulomb gauge
 at a finite periodic as well as twisted box of size $L^3$ and determine its form both in the derivative expansion and in the
  short-distance expansion to one-loop order in old-fashioned perturbation theory. In the Coulomb gauge, the Hamiltonian acting 
  on the reduced Hilbert space consists only of the physical degrees of freedom, all of the gauge degrees of freedom being eliminated. The momentum cutoff $\Lambda$ can be introduced consistently with Ward-Slavnov-Taylor identity \cite{2005AT,2006Niegawa} and this allows us to proceed to the straightforward short-distance expansion. 

In the next section, we give several preliminaries to the subsequent sections. In particular, we present position space expression 
of the Coulomb Green function (the inverse of the Laplacian) for the periodic sector and that for the twisted sectors. In section 
three, we consider the case of (periodic) QED for comparison with the pure non-abelian case and illustrate the derivative and the 
short-distance expansions. Section four contains main results of our paper.
 We deal with the non-abelian case to confirm the asymptotic freedom from the effective coupling 
constant and  to obtain the Uehling potential (see, for example, \cite{1995PS}) at the derivative expansion to one-loop order. 
The one-loop part of the short-distance  expansion begins with $\Lambda^2 /p^4$, 
which translates into $\Lambda^2 \int d^3z G(\bm{x}-\bm{z})G(\bm{z}-\bm{y})$ in position space, being non-singular at $\bm{x}=\bm{y}$. We determine the coefficient 
to one-loop order. 
The effect of the twist is seen through the phase factor of the Green function in the twisted sector by the Poisson resummation formula. In the final section, we briefly conclude our results in the bigger perspective.

\section{Some preliminaries}

\subsection{twisted boundary condition}
While it is not a main scope of this paper, pure non-abelian gauge theory permits twisted as well as periodic boundary condition due to the presence of the center of $SU(N)$ group. In this subsection, we will briefly recall this well-known fact and treat the cases of periodic boundary condition  and the twisted boundary conditions collectively.

Let $A_i (x,y,z)=\sum_{a}T^a A^a_i (x,y,z)$ be these spacial components of an $SU(N)$ gauge field, which  is Lie algebra valued. As we work on  Hamiltonian formalism, 
we will suppress time $t$ unless necessary. We adopt the twisted boundary condition of the following form:
\begin{align}
A_i (x,y,z)=PA_i (x+L,y,z)P^{-1}=QA_i (x,y+L,z)Q^{-1}=A_i (x,y,z+L),
\label{tbc}
\end{align}
where $P$ and $Q$ are the constant matrices which satisfy for $SU(N)$,
\begin{align}
	PQ=QP e^{\frac{2\pi i}{N}}.
\end{align}
An explicit representation for $P$ and $Q$ is 
\begin{align}
P=\alpha
\left(\begin{array}{cccccc}
0&1&0&  &&\\
  &0&1&0& &   \\
  &  &0 &1  &\ddots&\\
  &  &  &0&\ddots &0\\
  &  &&&\ddots&1\\
1&0&&& &0\\
\end{array}\right),\quad
Q=\beta
\left(\begin{array}{ccccc}
1&&&&\\
&e^{2\pi i/N}&&&\\
&&e^{4\pi i/N}&&\\
&&&\ddots&\\
&&&&e^{2\pi i(N-1)/N}\\
\end{array}\right).
\end{align}
Here $\alpha$ and $\beta$ are chosen so that $\det P =\det Q =1$ \cite{1982Wittenindex}. In the next subsection and the subsequent ones, we will work on an explicit solution to this  boundary condition in the case of $SU(2)$ only.

The extension to the explicit solution to the $SU(N)$ case ($N\geq 3$) is a straightforward eigenvalue problem in the linear algebra and will not be attempted here. In 't Hooft terminology, the twisted boundary condition \eqref{tbc} describes one of the three twisted sectors with a unit magnetic flux, the remaining two obtained by the cubic symmetry  of the box. There are another three sectors \eqref{tbc} having the magnetic fluxes in two different directions and one sector with the magnetic fluxes in all three directions.

\subsection{mode expansion and bracket notation}

In order to avoid using plane wave expressions in most places, we will adopt the bracket notation. Let $f(\bm{x})$ obey 
the twisted boundary condition labelled by $\bm{\lambda}$ and be expandable as Fourier series. Preparing the ket $|f \rangle$ and 
the bra vector $\langle \bm{x}|$ in the coordinate representation such that $\langle\bm{x}|\bm{x^{\prime}}\rangle=
\delta^{(3)}(\bm{x}-\bm{x^{\prime}})$ and therefore$\int d^3x' |\bm{x}'\rangle\langle\bm{x}'|=1$,
we write
\begin{align}
f(\bm{x})
=\langle\bm{x}|f\rangle
=\int d^3x^{\prime}\langle\bm{x}|\bm{x^{\prime}}\rangle f(\bm{x^{\prime}})
=\langle\bm{x}|\int d^3x'|\bm{x}'\rangle \langle\bm{x}'|f \rangle,
\label{notation_f(x)}
\end{align}
while
\begin{align}
f(\bm{x})
=\sum_{\bm{w}\in\mathbb{Z}^3 +\bm{\lambda}} C_{\bm{w}}^{(\bm{\lambda})} e^{\frac{2\pi i}{L}\bm{w}\cdot\bm{x}}
\equiv \sum_{\bm{w}\in\mathbb{Z}^3 +\bm{\lambda}} C_{\bm{w}}^{(\bm{\lambda})}
\langle\bm{x}|\bm{w}\rangle_{\bm{\lambda}}.
\label{notation_Fourier}
\end{align}
Here we have introduced the ket vector $|\bm{w}\rangle_{\bm{\lambda}}$ in the momentum representation in the $\bm{\lambda}$ 
twisted sector. In the twisted sector,
\begin{align}
\frac{1}{L^3}\int d^3x e^{\frac{2\pi i}{L}(\bm{w}-\bm{w}')\cdot\bm{x}}
=\frac{1}{L^3}\int d^3x {}_{\bm{\lambda}}\langle\bm{w}'|\bm{x}\rangle\langle\bm{x}|\bm{w}\rangle_{\bm{\lambda}}=\delta^{(3)}_{\bm{w},\bm{w}'},
\label{Kronecker_delta}
\end{align}
still holds, so that 
\begin{align}
C_{\bm{w}'}^{(\bm{\lambda})}
=\frac{1}{L^3}\int d^3x' {}_{\bm{\lambda}}\langle\bm{w}'|\bm{x}'\rangle\langle\bm{x}'|f\rangle.
\label{Fourier_mode}
\end{align}
Plugging this into \eqref{notation_Fourier} and comparing with  \eqref{notation_f(x)}, we obtain 
\begin{align}
\frac{1}{L^3}\sum_{\bm{w}\in\mathbb{Z}^3 +\bm{\lambda}}|\bm{w}\rangle_{\bm{\lambda}}{}_{\bm{\lambda}}\langle\bm{w}|=\bm{1}_{\bm{\lambda}}.
\label{relation_identity}
\end{align}
Here, we have denote by $\bm{1}_{\bm{\lambda}}$ the unit operater in the $\bm{\lambda}$ twisted sector.

Following the relativistic normalization seen in the standard textbook, we expand the gauge field $A_i^a (\bm{x})_{\bm{\lambda}^{(a)}}$ belonging to
the $\bm{\lambda}^{(a)}$ twisted sector and its canonical conjugate $\Pi_i^a (\bm{x})_{\bm{\lambda}^{(a)}}$ at $t=0$ as
\begin{align}
A_i^a (\bm{x})_{\bm{\lambda}^{(a)}}
&=\sum_{\bm{w}\in\mathbb{Z}^3 +\bm{\lambda}^{(a)}} 
\frac{1}{\sqrt{2\omega (\bm{w})L^3}} \biggr( \alpha_i^a (\bm{w}) \langle \bm{x}|\bm{w} \rangle_{\bm{\lambda}^{(a)}} + \text{h.c.} \biggl),\\
\Pi_i^a (\bm{x})_{\bm{\lambda}^{(a)}}
&=\sum_{\bm{w}\in\mathbb{Z}^3 +\bm{\lambda}^{(a)}} 
\sqrt{\frac{\omega (\bm{w})}{2 L^3}} \biggr( (-i) \alpha_i^a (\bm{w}) \langle \bm{x}|\bm{w}\rangle_{\bm{\lambda}^{(a)}} + \text{h.c.} \biggl),
\end{align}
where $\omega (\bm{w})=\frac{2\pi}{L}|\bm{w}|$. The solution to the twisted boundary condition \eqref{tbc} is
\begin{align}
	\bm{\lambda}^{(a=1)}=
	\left(\begin{array}{c}
		0\\
		1/2\\
		0
	\end{array}\right),\quad
	\bm{\lambda}^{(a=2)}=
	\left(\begin{array}{c}
		1/2\\
		1/2\\
		0
	\end{array}\right),\quad
	\bm{\lambda}^{(a=3)}=
	\left(\begin{array}{c}
		1/2\\
		0\\
		0
	\end{array}\right).
\end{align}
Here, the column vecters refer to the $x,y,z$ components. Quantization in the Coulomb gauge contains only the transverse part of the gauge fields: the physically relevant part of the oscillators is 
\begin{align}
\alpha_i^{\text{(tr)} a} (\bm{w}) 
\equiv P(\bm{w})_{ij} \alpha_j^a (\bm{w}),\quad
P(\bm{w})_{ij}=\delta_{ij}-\frac{w_i w_j}{\bm{w}\cdot\bm{w}}.
\end{align}
The canonical commutation relations are
\begin{align}
&\Bigr[\alpha_i^{\text{(tr)} a} (\bm{w}),\alpha_j^{\text{(tr)} b\dagger} (\bm{w}')\Bigl]
=\delta^{ab} \delta_{\bm{w},\bm{w}'} P(\bm{w})_{ij},\nonumber\\
&\Bigr[ \alpha_i^{\text{(tr)} a} (\bm{w}),\alpha_j^{\text{(tr)} b} (\bm{w}') \Bigl]
=\Bigr[ \alpha_i^{\text{(tr)} a\dagger} (\bm{w}),\alpha_j^{\text{(tr)} b\dagger} (\bm{w}') \Bigl]=0.
\end{align}


\subsection{Green function}

We will deal with the loop-corrected  Coulomb force potential in the subsequent sections. We list here 
the Green function of the Laplacian in the $\bm{\lambda}$ twisted sectior:
\begin{align}
	G^{\bm{\lambda}}(\bm{x}|\bm{x}')
	&=\bigr(\partial_i \partial_i\bigl)^{-1} \langle\bm{x}|\bm{x}'\rangle
	=\langle\bm{x}|\hat{\Delta}^{-1}|\bm{x}'\rangle
	=\biggr(\frac{L}{2\pi}\biggl)^2 \langle\bm{x}|\frac{1}{L^3}
	\biggr(\sum_{\substack{\bm{w}\in\mathbb{Z}^3 +\bm{\lambda} \\\bm{w}\neq\bm{0}}} \frac{(-1)}{\bm{w}\cdot\bm{w}} |\bm{w}\rangle_{\bm{\lambda}} {}_{\bm{\lambda}} 
	\langle\bm{w}|\biggl)|\bm{x}'\rangle\nonumber\\
	&=-\frac{1}{4\pi}\sum_{\bm{\ell}\in \mathbb{Z}^3}
	\frac{e^{2\pi i\bm{\lambda}\cdot \bm{\ell}}}{|\bm{x}-\bm{x}'+L\bm{\ell}|}.
\end{align}
The last expansion is obtained from the Poisson resummation formula, which we review in the appendix. In the case of the periodic sector $\bm{\lambda}=\bm{0}$, we obtain
\begin{align}
G^{\bm{\lambda}=\bm{0}}(\bm{x}|\bm{x}')
=\biggr(\frac{L}{2\pi}\biggl)^2 \langle\bm{x}|\frac{1}{L^3}
\biggr(\sum_{\substack{\bm{w}\in\mathbb{Z}^3\\\bm{w}\neq\bm{0}}} \frac{(-1)}{\bm{w}\cdot\bm{w}} |\bm{w}\rangle_{\bm{0}} {}_{\bm{0}}\langle\bm{w}| \biggl) 
|\bm{x}'\rangle
=-\frac{1}{4\pi}\sum_{\bm{\ell}\in \mathbb{Z}^3}\frac{1}{|\bm{x}-\bm{x}'+L\bm{\ell}|}.
\end{align}
This agrees with the Green function in the periodic box\footnote{The charge neutrality condition for the total source is required in the periodic box by the Gauss' law. This removes the zero-mode from our consideration.}.

The Green function $G^{\bm{\lambda}}(\bm{x}|\bm{x}')$ in the limit $L\rightarrow\infty$ does not depend on the twist $\bm{\lambda}$ 
and is  simply
\begin{align}
	G(\bm{x}|\bm{x}')
	=-\frac{1}{4\pi}\frac{1}{|\bm{x}-\bm{x}'|}.
\end{align}

\subsection{Coulomb gauge Hamiltonian}

Before presenting the Coulomb gauge Hamiltonian of non-abelian gauge theory which we work with in this paper, let us make a pedagogical outline of its derivation, starting from the operator formalism at the time-like axial gauge. Here, we closely follow the discussion of \cite{1980CL}. See also \cite{1981TDLee}. It is well-known that, in quantizing gauge theory in general, not all the equations of motion that holds at the classical level are maintained as operator equations. In the time-like axial gauge, the non-abelian analog of the Gauss law is not realized as an operator equation but instead is imposed on the state space as constraints. The quantization itself goes by the standard equal time commutation relations on positive definite Hilbert space. The transition from the time-like axial gauge with the Gauss law constraint to the Coulomb gauge is regarded as the change of coordinates from Cartesian to curvilinear ones in the infinite dimentional field space. The constraint gets eliminated by this procedure and the time components of the gauge field become dependent variables, giving rise to the non-abelian analog of the Coulomb potential. Another feature of this transition to the Coulomb gauge is that we must take care of the nontrivial Jacobian associated with this transformation, which is nothing but the Faddeev-Popov determinant. Finally, by a similarity transformation, we obtain the Coulomb gauge Hamiltonian acting on the reduced Hilbert space consistenting of transverse physical degrees of freedom only.

Here we just list the Hamiltonian
\begin{align}
H&=\frac{1}{2} \sum_{a} \int d^3x \biggr( \mathcal{J}^{-1} \Pi_i^{\text{(tr)}a}\mathcal{J}\Pi_i^{\text{(tr)}a} + B_i^{\text{(tr)}a} B_i^{\text{(tr)}a} \biggl)+H_{\text{Coul}},
\label{Hamiltonian_total}\\
H_{\text{Coul}}&=\frac{g_0^2}{2} \sum_{a,b} \int d^3x d^3x'\mathcal{J}^{-1}\rho^a(\bm{x}) 
\langle\bm{x}|\Bigr( (\partial_i D_i)^{-1} (-\partial^2 ) (\partial_j D_j)^{-1} \Bigl)^{ab} |\bm{x}'\rangle \mathcal{J}\rho^b(\bm{x}'),
\label{Hamiltonian_Coul}
\end{align}
where $\Pi_i^{\text{(tr)}a}$, $B_i^{\text{(tr)}a}$, $D_i$, $\rho^a$ and $\mathcal{J}$ are respectively the conjugate momentum of the transverse gauge field $A_i^{\text{(tr)}a}$, the transverse magnetic field, covariant derivative, the charge density and the Faddeev-Popov determinant $\mathcal{J}=\det (\partial_i D_i)$. As we are in the Coulomb gauge, only the transverse parts of the gauge field $A_i$ (and its canonical conjugate $\Pi_i$) contribute to the Hamiltonian. For simplicity, we will omit the symbol (tr) in the following discussion. We expand the operater $((\partial_i D_i)^{-1} (-\partial^2 ) (\partial_j D_j)^{-1})^{ab}$ in the Coulomb potential part $H_{\text{Coul}}$ of the Hamiltonian by the coupling constant $g_0$ as follows:
\begin{align}
&\langle\bm{x}|\Bigr( (\partial_i D_i)^{-1} (-\partial^2 ) (\partial_j D_j)^{-1} \Bigl)^{ab} |\bm{x}'\rangle\nonumber\\
&=\langle\bm{x}| \Bigr( -\bigr( \hat{\Delta}^{-1} \bigl)^{ab} 
+2g_0 \bigr( \hat{\Delta}^{-1}\hat{\Omega}\hat{\Delta}^{-1} \bigl)^{ab}
-3g_0^2 \bigr( \hat{\Delta}^{-1}\hat{\Omega} \hat{\Delta}^{-1}\hat{\Omega}\hat{\Delta}^{-1} \bigl)^{ab} + \mathcal{O}(g_0^3)\Bigl) |\bm{x}'\rangle,
\end{align}
where $(\hat{\Delta}^{-1})^{ab}$, $\hat{\Omega}^{ab}$ are the operators respectively represented as
\begin{align}
\langle\bm{x}|\bigr(\hat{\Delta}^{-1}\bigl)^{ab}|\bm{x}'\rangle
&=\delta^{ab}G^{\bm{\lambda^{(b)}}}(\bm{x}|\bm{x}'),\\
\langle\bm{x}|\hat{\Omega}^{ac}|\bm{x}'\rangle
&=\delta^{(3)}(\bm{x}-\bm{x}')\epsilon^{abc}A_i^b(\bm{x}')\frac{\partial}{\partial x'^i}.
\end{align}
Here, the totally antisymmetric tensor $\epsilon^{abc}$ is the structure constant of the gauge group $SU(2)$. In the Hamiltonian, we have included two of the classical external source terms in the charge density $\rho^a(\bm{x})$
\begin{align}
\rho^a(\bm{x})
&=g_0\epsilon^{abc} A_i^b(\bm{x})\Pi_i^c(\bm{x})
+\rho^a_{1,\text{ex}}(\bm{x})+\rho^a_{2,\text{ex}}(\bm{x}),\nonumber\\
\rho^a_{1,2,\text{ex}}(\bm{x})
&=\sum_{\bm{w}\in\mathbb{Z}^3 +\bm{\lambda}^{(a)}}
\langle\bm{x}|\bm{w}\rangle_{\bm{\lambda}^{(a)}} \tilde{\rho}^a_{1,2,\text{ex}}(\bm{w}).
\label{source}
\end{align}
The two  delta finction sources localized at $\bm{x}=\bm{x}_1,\bm{x}_2$ are respectively represented as
\begin{align}
\tilde{\rho}^a_{1,2,\text{ex}}(\bm{w})
=q^a_{1,2}\frac{1}{L^3}{}_{\bm{\lambda}^{(a)}}\langle\bm{w}|\bm{x}_{1,2}\rangle,
\end{align}
so
\begin{align}
\rho^a_{1,2,\text{ex}}(\bm{x})
=q^a_{1,2}\frac{1}{L^3}\sum_{\bm{w}\in\mathbb{Z}^3 +\bm{\lambda}^{(a)}}
\langle\bm{x}|\bm{w}\rangle_{\bm{\lambda}^{(a)}} {}_{\bm{\lambda}^{(a)}}\langle \bm{w}|\bm{x}_{1,2}\rangle
= q^a_{1,2}\delta^{(3)}(\bm{x}-\bm{x}_{1,2}),
\end{align}
in the $\bm{\lambda}$ twisted sector.

We will be interested in the part of the vacuum energy which depends linearly upon 
both $\rho^a_{1,\text{ex}}(\bm{x})$ and $\rho^a_{2,\text{ex}}(\bm{x})$. Clearly, at the lowest classical level,
\begin{align}
E_{\text{tw}}
&=-g_0^2 \sum_{a} \int d^3xd^3x' \rho^a_{1,\text{ex}}(\bm{x}) G^{\bm{\lambda}^{(a)}}(\bm{x}|\bm{x}')\delta^{ab} \rho^b_{2,\text{ex}}(\bm{x}')\nonumber\\
&=-g_0^2 \sum_{a} q_1^a q_2^a G^{\bm{\lambda}^{(a)}}(\bm{x}_1 |\bm{x}_2).
\end{align}

\section{Case of QED}
In this section, we obtain, to one-loop order, the interaction energy between the two external static sources with charges $q_1$ and $q_2$ for QED in the periodic box of size $L^3$ by old-fashioned perturbation theory well-known in quantum mechanics. We will confirm the UV divergence and the renormalization of the coupling constant and the Uehling potential in QED for the massless fermions at finite volume.

Let us first denote the free part and the interaction part of the Coulomb gauge Hamiltonian $H$ by $H^{(0)}$ and $H_{\text{int}}$ respectively:
\begin{align}
	H_{\text{int}} \equiv H-H^{(0)}.
\end{align}
The massless fermions are expanded as 
\begin{align}
\psi(\bm{x})
&=\sum_{\bm{w}\in\mathbb{Z}^3}
\frac{1}{\sqrt{2\omega(\bm{w})L^3}} \Bigr( d^s(\bm{w})u^s(\bm{w}) e^{i\frac{2\pi}{L}\bm{w}\cdot \bm{x}}+ {b^s}^\dagger(\bm{w})v^s(\bm{w}) e^{-i\frac{2\pi}{L}\bm{w}\cdot \bm{x}} \Bigl),\\
\bar{\psi}(\bm{x})
&=\sum_{\bm{w}\in\mathbb{Z}^3}
\frac{1}{\sqrt{2\omega(\bm{w})L^3}} \Bigr( b^s(\bm{w})\bar{v}^s(\bm{w}) e^{i\frac{2\pi}{L}\bm{w}\cdot
\bm{x}}+ {d^s}^\dagger(\bm{w})\bar{u}^s(\ \bm{w}) e^{-i\frac{2\pi}{L}\bm{w}\cdot \bm{x}} \Bigl).
\end{align}
The Hamiltonian in this section includes the kinetic term for the fermions and their gauge interactions. The charge density is
\begin{align}
	\rho (\bm{x})=\rho_{1,\text{ex}}(\bm{x})+\rho_{2,\text{ex}}(\bm{x})+\bar{\psi}\gamma^0 \psi.
\end{align}

The eigenstates and the eigenvalues for the Hamiltonian of $H$ and the free part of the Hamiltonian $H^{(0)}$ are respectively
denoted by
\begin{align}
	H|N_{\alpha};N_{d},N_{b}\rangle&=E_{N_{\alpha};N_{d},N_{b}}|N_{\alpha};N_{d},N_{b}\rangle,\\
	H^{(0)}|N_{\alpha};N_{d},N_{b}\rangle&=E^{(0)}_{N_{\alpha};N_{d},N_{b}}|N_{\alpha};N_{d},N_{b}\rangle,\\
	H^{(0)}|0;0,0\rangle&=E^{(0)}_{0;0,0}|0;0,0\rangle =0,
\end{align}
where $N_\alpha$ is the number of bosons and $N_d$ and $N_b$ are  respecgtively the number of fermions and that of antifermions. As we are interested in the interaction energy between the two external sources, we ignore
the zero point oscillation and set $E^{(0)}_{0;0,0}=0$\footnote{In this paper, we do not estimate the contributions 
coming from the zero momentum modes and possible infrared divergences associated with them.}. 

As in quantum mechanics, the perturbative expansion of $E(r_{12})$ goes as
\begin{align}
E(r_{12})
&=E^{(\text{i})}(r_{12})+E^{(\text{ii})}(r_{12})+\cdots,\\
E^{(\text{i})}(r_{12})
&=\langle 0|H_{\text{int}}(r_{12})|0\rangle,\\
E^{(\text{ii})}(r_{12})
&=\sum_{\substack{ N_{\alpha};N_{d},N_{b} \\ (N_{\alpha};N_{d},N_{b})\neq (0;0,0)}}\frac{|\langle 0|H_{\text{int}}(r_{12})|N_{\alpha};N_{d},N_{b}\rangle|^2}{-E^{(0)}_{N_{\alpha};N_{d},N_{b}}},\cdots.
\end{align}
After some calculation which we omit presenting here (it is a routine), we obtain  the leading order $E^{(\text{i})}(r_{12})$ and 
the second order $E^{(\text{ii})}(r_{12})$  corrections respectively given by
\begin{align}
E^{(\text{i})}(r_{12})
=\frac{g_0^2 q_1 q_2}{4\pi^2 L}\sum_{\substack{\bm{n}\in\mathbb{Z}^3\\ \bm{n}\neq\bm{0}}}\frac{e^{i\frac{2\pi}{L}\bm{n}\cdot(\bm{x}_2-\bm{x}_1)}}{\bm{n}\cdot\bm{n}}
=\frac{1}{4\pi}\sum_{\bm{\ell}\in\mathbb{Z}^3} \frac{g_0^2 q_1 q_2}{|\bm{x}_1-\bm{x}_2+\bm{\ell}L|},
\end{align}
exploiting the Poisson resummation formula, and
\begin{align}
E^{(\text{ii})}(r_{12})
&=\frac{g_0^2 q_1 q_2}{4\pi^2 L}\sum_{\substack{\bm{n}\in\mathbb{Z}^3\\ \bm{n}\neq\bm{0}}}\frac{e^{i\frac{2\pi}{L}\bm{n}\cdot(\bm{x}_2-\bm{x}_1)}}{\bm{n}\cdot\bm{n}} \bigr[ g_0^2 \delta(\bm{n}) \bigl],
\label{QED_second}\\
\delta(\bm{n})
&=-\frac{1}{4\pi^3} \frac{1}{\bm{n}\cdot\bm{n}}\sum_{\substack{\bm{m}\in\mathbb{Z}^3\\ \bm{m}\neq\bm{0}}} \frac{1}{|\bm{m}|+|\bm{n}+\bm{m}|} \biggr( 1-\frac{\bm{m}\cdot(\bm{n}+\bm{m})}{|\bm{m}||\bm{n}+\bm{m}|} \biggl)\nonumber.
\end{align}

\subsection{derivative expansion }
Let us first consider the derivative expansion, which will be valid at the distance comparable to the size of the box, to evaluate the first quantum correction \eqref{QED_second} to Coulomb potential in perturbation theory. The derivative expansion corresponds to the triple Taylor expansion with respect to $n^i$ ($i=1,2,3$). We obtain
\begin{align}
&\delta (\bm{n})
=-\frac{1}{4\pi^3} \frac{1}{\bm{n}\cdot\bm{n}} f(\bm{n})\nonumber\\
&=-\frac{1}{4\pi^3} \frac{1}{\bm{n}\cdot\bm{n}} \biggl( f(\bm{0}) 
+ \frac{1}{2!} \frac{\partial^2 f}{\partial n^i \partial n^j} (\bm{0}) n^i n^j 
+ \frac{1}{4!} \frac{\partial^4 f}{\partial n^i \partial n^j \partial n^k \partial n^\ell} (\bm{0}) n^i n^j n^k n^\ell
+\cdots \biggr),
\label{QEDdelta}
\end{align}
where we  have omitted the terms odd in $n^i$ as they cancel upon taking the summation over $n^i$. Using the symmetry of cubic lattice, the above expansion \eqref{QEDdelta} is written as
\begin{align}
\delta (\bm{n})
=-\frac{1}{4\pi^3} \sum_{\substack{\bm{m}\in\mathbb{Z}^3\\ \bm{m}\neq \bm{0}}} 
\biggr[\frac{1}{6|\bm{m}|^3}+\frac{1}{120}\frac{|\bm{n}|^2}{|\bm{m}|^5}
+\mathcal{O}(|\bm{n}|^4) \biggl].
\end{align}
Coming back to \eqref{QED_second} and using the Poisson resummation formula, we obtain
\begin{align}
E^{\text{(ii)}}(r_{12}) =
&\frac{1}{4\pi}\sum_{\bm{n}\in\mathbb{Z}^3} \frac{g_0^2 q_1 q_2}{|\bm{x}_1-\bm{x}_2+\bm{n}L|} \biggl(-\frac{g_0^2}{6\pi^2}\biggr) \frac{1}{4\pi} \sum_{\substack{\bm{m}\in\mathbb{Z}^3\\ \bm{m}\neq \bm{0}}}\frac{1}{|\bm{m}|^3}\nonumber\\
&-\frac{g_0^4 q_1 q_2}{(4\pi)^2}\frac{1}{15}\frac{1}{(2\pi)^3} 
\biggl( L^2\sum_{\bm{n}\in\mathbb{Z}^3} \delta^{(3)}(\bm{x}_1-\bm{x}_2+\bm{n}L) -\frac{1}{L} \biggr)
\sum_{\substack{\bm{m}\in\mathbb{Z}^3\\ \bm{m}\neq \bm{0}}} \frac{1}{|\bm{m}|^5}\nonumber\\
&+\text{(higher orders in the derivative expansion)} .
\end{align}
The higher orders will give (gaussian) width to the delta function potential. This expression is still at  finite volume $L^3$.

Taking the large volume, up to the second order in perturbation theory, we obtain  the derivative expansion of
the loop-corrected Coulomb potential as 
the interaction energy between the two external charges $q_1$ and $q_2$:
\begin{align}
E(r_{12})
=\frac{1}{4\pi}\frac{q_1 q_2}{r_{12}} g_0^2 \biggl(1-\frac{g_0^2}{6\pi^2}\ln (\Lambda L) \biggr)
-\frac{g_0^4 q_1 q_2}{(4\pi)^2}\frac{1}{15} L^2 \biggr(1-\frac{1}{(\Lambda L)^2}\biggl) \delta^{(3)}(\bm{x}_1-\bm{x}_2),
\label{QEDLinfinity}
\end{align}
where $\Lambda$ is the UV cutoff. The first term  derives the positive $\beta$-function of QED at one-loop while the second term is  the Uehling potential. 
Here, when we define the renormalized coupling constant $g_L$ in the box $L^3$:
\begin{align}
g_L^2=g_0^2 \biggl(1-\frac{g_0^2}{6\pi^2}\ln (\Lambda L) \biggr),
\end{align}
$g_L^2$ is written in $g_0^2$:
\begin{align*}
g_0^2=g_L^2 \biggl(1+\frac{g_L^2}{6\pi^2}\ln (\Lambda L) \biggr)+\mathcal{O}(g_0^6).
\end{align*}
By substituting this into \eqref{QEDLinfinity}, we can also write $E(r_{12})$ as 
\begin{align}
E(r_{12})
\simeq\frac{1}{4\pi}\frac{q_1 q_2}{r_{12}} g_L^2
-\frac{g_L^4 q_1 q_2}{(4\pi)^2}\frac{1}{15} L^2 \biggr(1-\frac{1}{(\Lambda L)^2}\biggl) \delta^{(3)}(\bm{x}_1-\bm{x}_2),
\end{align}
at the order $g_L^4$.

\subsection{expansion at  short-distance}
Let us now probe the opposite limit to the last subsection. We will evaluate the interaction energy by the short-distance $r_{12}\ll L$ expansion. In this expansion, $E(r_{12})$ is expanded in $\Lambda /p$. Here, we take the limit $L\rightarrow\infty$ from the beginning.

We obtain
\begin{align}
	E(r_{12})
	&=\frac{g_0^2 q_1 q_2}{(2\pi)^3}\int d^3p \frac{1}{\bm{p}\cdot\bm{p}} e^{i\bm{p}\cdot (\bm{x}_1-\bm{x}_2)}\biggr[1+ g_0^2 \delta (\bm{p}) \biggl],\\
	\delta (\bm{p})
	&=-\frac{1}{4\pi^3}\frac{1}{\bm{p}\cdot\bm{p}} \int d^3k \frac{1}{|\bm{k}|+|\bm{p}+\bm{k}|} \biggl(1-\frac{\bm{k}\cdot (\bm{p}+\bm{k})}{|\bm{k}||\bm{p}+\bm{k}|}\biggr).
\end{align}
Using the polar coordinates, $\delta (\bm{p})$ is further converted as
\begin{align}
\delta (\bm{p})
&=-\frac{1}{2\pi^2}\frac{1}{\bm{p}\cdot\bm{p}}\int^{\Lambda} k^2 dk \int_{0}^{\pi}d\theta \frac{\sin\theta}{k+\sqrt{k^2+p^2+2 k p \cos\theta}} \biggl( 1- \frac{k+p\cos\theta}{\sqrt{k^2+p^2+2 k p \cos\theta}} \biggr)\nonumber\\
&=-\frac{1}{2\pi^2} \frac{1}{\bm{p}\cdot\bm{p}} \int_{0}^{\pi}d\theta \sin\theta I(\Lambda,p,\theta),\label{bm{p}}\\
I(\Lambda,p,\theta)
&=\int^{\Lambda} k^2 dk \frac{1}{k+\sqrt{k^2+p^2+2 k p \cos\theta}} \biggl( 1- \frac{k+p\cos\theta}{\sqrt{k^2+p^2+2 k p \cos\theta}} \biggr).
\end{align}
Expanding $I(\Lambda,p,\theta)/\Lambda^2$ in $\frac{\Lambda}{p}$, we obtain
\begin{align}
I(\Lambda,p,\theta)=\frac{1-\cos\theta}{3}\frac{\Lambda^3}{p}+\frac{-1+\cos^2 \theta}{2}\frac{\Lambda^4}{p^2}+\mathcal{O}\biggl(\frac{\Lambda^5}{p^3}\biggr).
\end{align}
\eqref{bm{p}} becomes
\begin{align}
\delta (\bm{p})
&=-\frac{1}{2\pi^2}\int_{0}^{\pi}d\theta \sin\theta \biggl(
\frac{1-\cos\theta}{3}\biggl(\frac{\Lambda}{p}\biggr)^3+\frac{-1+\cos^2 \theta}{2}\biggl(\frac{\Lambda}{p}\biggr)^4+\mathcal{O}\biggl(\frac{\Lambda^5}{p^5}\biggr)\biggr)\nonumber\\
&=-\frac{1}{3\pi^2}\biggl(
\biggl(\frac{\Lambda}{p}\biggr)^3-\biggl(\frac{\Lambda}{p}\biggr)^4+\mathcal{O}\biggl(\frac{\Lambda^5}{p^5}\biggr)\biggr).
\end{align}
The loop-corrected potential  as the interaction energy $E(r_{12})$ at  short-distance expansion is thus given by
\begin{align}
E(r_{12})
&=\frac{g_0^2 q_1 q_2}{(2\pi)^3}\int d^3p
e^{i\bm{p}\cdot (\bm{x}_2-\bm{x}_2)}\biggr[\frac{1}{p^2} -\frac{g_0^2}{3\pi^2}\biggl( \frac{\Lambda^3}{p^5}-\frac{\Lambda^4}{p^6}+\mathcal{O}\biggl(\frac{\Lambda^5}{p^7}\biggr)\biggr)\biggl].
\end{align}

\section{Static force potential in pure non-abelian gauge theory at a periodic and twisted box}

Let us now turn to the case of pure non-abelian gauge theory at a twisted or periodic finite box of size $L^3$. Unlike the periodic one, the zero-mode is not present in the twisted sector. We manage to treat both cases at once in the notation in what follows.

Following the method in the abelian case, we will compute the interaction energy  $E(r_{12})$ between the two external static sources of charge $q_1^a$ and  $q_2^a$ in old-fashioned perturbation theory. Up to the second order, it reads
$E(r_{12})=E^{(\text{i})}(r_{12})+E^{(\text{ii})}(r_{12})$:
\begin{align}
E^{(\text{i})}(r_{12})
&=\sum_{a} \frac{q_1^a q_2^a}{4\pi^2 L}\sum_{\substack{\bm{n}\in\mathbb{Z}^3 +\bm{\lambda}^{(a)} \\ \bm{n}\neq\bm{0}}}
\langle \bm{x}_1 |\bm{n}\rangle_{\bm{\lambda}^{(a)}}
\frac{1}{\bm{n}\cdot\bm{n}} \biggl( g_0^2+g_0^4 \delta'_{\bm{\lambda}^{(a)}} (\bm{n}) \biggr) {}_{\bm{\lambda}^{(a)}} \langle \bm{n}|\bm{x}_2 \rangle,\\
E^{(\text{ii})}(r_{12})
&=\sum_{a} \frac{q_1^a q_2^a}{4\pi^2 L}\sum_{\substack{\bm{n}\in\mathbb{Z}^3 +\bm{\lambda}^{(a)} \\ \bm{n}\neq\bm{0}}}
\langle \bm{x}_1 |\bm{n}\rangle_{\bm{\lambda}^{(a)}}
\frac{1}{\bm{n}\cdot\bm{n}} \biggl(g_0^4 \delta''_{\bm{\lambda}^{(a)}}(\bm{n}) \biggr) {}_{\bm{\lambda}^{(a)}}
\langle \bm{n}|\bm{x}_2 \rangle,
\end{align}
where 
\begin{align}
\delta'_{\bm{\lambda}^{(a)}}(\bm{n})
&=\frac{3}{16\pi^3} \frac{1}{\bm{n}\cdot\bm{n}} \sum_{c\neq a} \sum_{\substack{\bm{m}\in\mathbb{Z}^3 +\bm{\lambda}^{(c)} \\ \bm{m}\neq\bm{0}}} \frac{(n+m)_i P(\bm{m})_{ij} n_j}{|\bm{n}+\bm{m}|^2 |\bm{m}|},\\
\delta''_{\bm{\lambda}^{(a)}}(\bm{n})
&=-\frac{1}{32\pi^3} \frac{1}{\bm{n}\cdot\bm{n}} \sum_{c\neq a} \sum_{\substack{\bm{m}\in\mathbb{Z}^3 +\bm{\lambda}^{(c)} \\\bm{m}\neq\bm{0}}} \frac{(|\bm{m}|-|\bm{n}-\bm{m}|)^2}{|\bm{m}|+|\bm{n}-\bm{m}|} \frac{P(\bm{m})_{ij} P(\bm{n-m})_{ij}}{|\bm{m}||\bm{n}-\bm{m}|}.
\end{align}
Here we have used 
\begin{align}
	\sum_{b,c} \epsilon^{bac} \epsilon^{bdc} f^{(c)}
	= \sum_{c} (\delta^{ad}\delta^{cc}-\delta^{ac}\delta^{dc}) f^{(c)}
	= \delta^{ad} \sum_{c} (\delta^{cc}-\delta^{ac}) f^{(c)}
	=\delta^{ad} \sum_{c\neq a} f^{(c)}.
\end{align}
In the case of $SU(N)$, we need only to replace $\epsilon^{bac}$ by the structure constant. (The above derivation, of course, 
does not hold.) As for $E^{(\text{i})}(r_{12})$, recall that $H_{{\rm Coul}}$ in the non-abelian case in \eqref{Hamiltonian_Coul} takes a non-local expression in the position space and generates infinite series of gauge fields in coupling constant. It starts contributing already at the first order in perturbation theory. We have a result in $E^{(\text{ii})}(r_{12})$ similar to that in the abelian case, replacing the fermionic intermediate states by the bosonic ones. Again, we have omitted presenting our calculation.

The interaction energy in the twisted box $L^3$ up to $g_0^4$ reads
\begin{align}
E(r_{12})=\sum_{a} \frac{g_0^2 q_1^a q_2^a}{4\pi^2 L}\sum_{\substack{\bm{n}\in\mathbb{Z}^3 +\bm{\lambda}^{(a)} \\ \bm{n}\neq\bm{0}}}
\langle \bm{x}_1 |\bm{n}\rangle_{\bm{\lambda}^{(a)}}
\frac{1}{\bm{n}\cdot\bm{n}} \biggl[ 1+g_0^2 \bigg( \delta'_{\bm{\lambda}^{(a)}} (\bm{n})+ 
\delta''_{\bm{\lambda}^{(a)}} (\bm{n}) \bigg) \biggr] {}_{\bm{\lambda}^{(a)}} \langle \bm{n}|\bm{x}_2 \rangle.
\label{E12}
\end{align}
The case of the periodic sector can be read off  from this expression by setting $\bm{\lambda}^{(a)} =0$.
The presence of $\delta'_{\bm{\lambda}} (\bm{n})$ from the contribution $E^{(\text{i})}(r_{12})$ is a unique feature of non-abelian gauge theory responsible for the asymptotic freedom.

\subsection{derivative expansion}

To evaluate the above result, we expand the quantum corrections $\delta'_{\bm{\lambda}^{(a)}}(\bm{n})+\delta''_{\bm{\lambda}^{(a)}}(\bm{n})$ in $n^i$ as in the abelian case:
\begin{align}
&\delta'_{\bm{\lambda}^{(a)}}(\bm{n})+\delta''_{\bm{\lambda}^{(a)}}(\bm{n})\nonumber\\
&=\frac{1}{32\pi^3} \sum_{c\neq a} 
\sum_{\substack{\bm{m}\in\mathbb{Z}^3 +\bm{\lambda}^{(c)}\\\bm{m}\neq\bm{0}}}
\biggl[ \frac{1}{2!}\biggl( \frac{12}{|\bm{m}|^3}-\frac{14 (\bm{m}\cdot\bm{n})^2}{|\bm{m}|^5 |\bm{n}|^2} \biggr)
+\frac{1}{3!} \biggl( -\frac{66 (\bm{m}\cdot\bm{n})}{|\bm{m}|^5} +\frac{57 (\bm{m}\cdot\bm{n})^3}{|\bm{m}|^7 |\bm{n}|^2} \biggr)\nonumber\\
&+\frac{1}{4!} \biggl( -\frac{150|\bm{n}|^2}{|\bm{m}|^5}+\frac{822 (\bm{m}\cdot\bm{n})^2}{|\bm{m}|^7}-\frac{714 (\bm{m}\cdot\bm{n})^4}{|\bm{m}|^9 |\bm{n}|^2}\biggr)
+(\text{higher orders of }n^i)
\biggr].
\end{align}
In the twisted sectors, namely $\bm{\lambda}\neq\bm{0}$, we can not use the cubic symmetry to evaluate the parts involving $\bm{m}\cdot\bm{n}$ in the box $L^3$. Here we consider the case of the periodic box, namely the $\bm{\lambda}=\bm{0}$ case only. Due to the cubic symmetry, the quantum corrections reduce to
\begin{align}
\delta'_{\bm{\lambda}=\bm{0}}(\bm{n})+\delta''_{\bm{\lambda}=\bm{0}}(\bm{n})
=\frac{1}{16\pi^3} \sum_{\substack{\bm{m}\in\mathbb{Z}^3 \\\bm{m}\neq\bm{0}}} \biggl[
\frac{11}{3}\frac{1}{|\bm{m}|^3}
-\frac{47}{60} \frac{|\bm{n}|^2}{|\bm{m}|^5}
+\mathcal{O}(|\bm{n}|^4)
\biggr].
\end{align}
We obtain the interaction energy at the periodic box:
\begin{align}
E(r_{12})=
&\frac{1}{4\pi} \sum_{a} \sum_{\bm{n}\in\mathbb{Z}^3} \frac{g_0^2 q_1^a q_2^a }{|\bm{x}_1-\bm{x}_2+\bm{n}L|} \biggl[ 1+ g_0^2 \frac{11}{12\pi^2} \frac{1}{4\pi} \sum_{\substack{\bm{m}\in\mathbb{Z}^3 \\\bm{m}\neq\bm{0}}} \frac{6}{|\bm{m}|^3} \biggr]\nonumber\\
&-\sum_{a} \frac{g_0^4 q_1^a q_2^a}{(4\pi)^2}\frac{47}{30} \frac{1}{(2\pi)^3}
\biggl( L^2\sum_{\bm{n}\in\mathbb{Z}^3} \delta^{(3)}(\bm{x}_1-\bm{x}_2+\bm{n}L) -\frac{1}{L} \biggr) \sum_{\substack{\bm{m}\in\mathbb{Z}^3 \\\bm{m}\neq\bm{0}}} \frac{1}{|\bm{m}|^5}\nonumber\\
&+\text{(higher orders in the derivative expansion)}.
\end{align}

For large $L$, the static force potential is evaluated as follows up to the second order in perturbation theory,
\begin{align}
E(r_{12})
=\sum_{a} \frac{q_1^a q_2^a}{4\pi r_{12}}g_0^2 \biggl( 1+g_0^2 \frac{11}{12\pi^2} \ln (\Lambda L) \biggr) 
-\sum_{a} \frac{g_0^4 q_1^a q_2^a}{(4\pi)^2}\frac{47}{30} L^2 \biggl( 1-\frac{1}{(\Lambda L)^2} \biggr) \delta^{(3)} (\bm{x}_1-\bm{x}_2),
\label{QCDLinfinity}
\end{align}
where $\Lambda$ is the UV cutoff. The first term of \eqref{QCDLinfinity} derives the negative $\beta$-function with the corect numerical coefficient\footnote{We regard this as support for the validity of the use of momentum cutoff to one-loop order. See \cite{1981TDLee} for further discussion on renormalization and renormalization group in this treatment.} and the second term is nothing but the Uehling potential.

Also, by defining the renormalized coupling constant $g_L$ as
\begin{align*}
g_L^2= g_0^2 \biggl( 1+g_0^2 \frac{11}{12\pi^2} \ln (\Lambda L) \biggr),
\end{align*}
we can also write $E(r_{12})$ as
\begin{align}
E(r_{12})
=\sum_{a} \frac{q_1^a q_2^a}{4\pi r_{12}}g_L^2 
-\sum_{a} \frac{g_L^4 q_1^a q_2^a}{(4\pi)^2}\frac{47}{30} L^2 \biggl( 1-\frac{1}{(\Lambda L)^2} \biggr) \delta^{(3)} (\bm{x}_1-\bm{x}_2),
\end{align}
at the order $g_L^4$.

\subsection{expansion at  short-distance}

Let us carry out the short-distance expansion as in the abelian case. Taking $L\rightarrow\infty$ limit, we obtain
\begin{align}
E(r_{12})=\sum_{a} \frac{g_0^2 q_1^a q_2^a}{(2\pi)^3} 
\int d^3p \frac{1}{\bm{p}\cdot\bm{p}} e^{i\bm{p}\cdot (\bm{x}_1-\bm{x}_2)} 
\biggl[ 1+g_0^2 \bigg( \delta' (\bm{p})+\delta'' (\bm{p}) \bigg) \biggr],
\end{align}
where
\begin{align}
\delta'(\bm{p})
&=\frac{3}{8\pi^3} \frac{1}{\bm{p}\cdot\bm{p}} \int d^3k \frac{(p+k)_i P(\bm{k})_{ij} p_j}{|\bm{p}+\bm{k}|^2 |\bm{k}|},\\
\delta''(\bm{p})
&=-\frac{1}{16\pi^3} \frac{1}{\bm{p}\cdot\bm{p}} \int d^3k \frac{(|\bm{k}|-|\bm{p}-\bm{k}|)^2}{|\bm{k}|+|\bm{p}-\bm{k}|} \frac{P(\bm{k})_{ij} P(\bm{p-k})_{ij}}{|\bm{k}||\bm{p}-\bm{k}|}.
\end{align}
Using the polar coordinates, we obtain
\begin{align}
	\delta'(\bm{p})
	=&\frac{3}{4 \pi^2} \frac{1}{\bm{p}\cdot\bm{p}}\int^{\Lambda}dk \int^{\pi}_{0} d\theta \sin\theta \frac{k p^2 (1-\cos^2 \theta)}{p^2 +k^2 +2pk \cos \theta}\nonumber\\
	=&\frac{3}{4 \pi^2} \frac{1}{\bm{p}\cdot\bm{p}} \int^{\pi}_{0} d\theta \sin\theta I'(\Lambda,p,\theta),\\
	\delta''(\bm{p})
	=&-\frac{1}{8\pi^2} \frac{1}{\bm{p}\cdot\bm{p}} \int^{\Lambda} \int_{0}^{\pi} d\theta \sin\theta \frac{k}{k+\sqrt{p^2+k^2-2pk\cos\theta}}\nonumber\\
	&\cdot \frac{(k-\sqrt{p^2+k^2-2pk\cos\theta})^2}{\sqrt{p^2+k^2-2pk\cos\theta}} \biggl( 1+\frac{(k-p\cos\theta)^2}{p^2+k^2-2pk\cos\theta} \biggr)\nonumber\\
	=&-\frac{1}{8 \pi^2} \frac{1}{\bm{p}\cdot\bm{p}} \int^{\pi}_{0} d\theta \sin\theta I''(\Lambda,p,\theta).
\end{align}
Expanding in $I'(\Lambda,p,\theta),I''(\Lambda,p,\theta)$ in $\Lambda$:
\begin{align}
	I'(\Lambda,p,\theta)=\frac{1-\cos^2 \theta}{2} \Lambda^2 +\cdots,\\
	I''(\Lambda,p,\theta)=\frac{1+\cos^2 \theta}{2} \Lambda^2 +\cdots,
\end{align}
we obtain
\begin{align}
E(r_{12})=\sum_{a} \frac{g_0^2 q_1^a q_2^a}{(2\pi)^3} \int d^3p  e^{i\bm{p}\cdot (\bm{x}_1-\bm{x}_2)} 
\biggl[ \frac{1}{p^2}+g_0^2 \bigg( \frac{1}{3\pi^2}\frac{\Lambda^2}{p^4}+\mathcal{O}\biggl(\frac{\Lambda^3}{p^5}\biggr) \bigg) \biggr].
\end{align}

We note that the $1/p^4$-term is written as
\begin{align}
\int d^3 z G(\bm{x}_1|\bm{z}) G(\bm{z}|\bm{x}_2)
=\int \frac{d^3 p}{(2\pi)^3}\frac{1}{p^4} e^{i\bm{p}\cdot (\bm{x}_1-\bm{x}_2 )},
\end{align}
which is non-singular for the position at $\bm{x}_1=\bm{x}_2$:
\begin{align}
\int d^3 z G(\bm{x}|\bm{z}) G(\bm{z}|\bm{x}')
=\biggl( \frac{-1}{4\pi} \biggr)^2 \int d^3 z \frac{1}{|\bm{x}-\bm{z}|} \frac{1}{|\bm{z}-\bm{x}'|}.
\label{2Green}
\end{align}
Here the integral in the right-hand side of \eqref{2Green} is divergent in the large $L$ limit, however the divergence depends on the volume of the box $L^3$ only and, except this divergence, there is no divergence. In terms of the Green functions, the interaction energy $E(r_{12})$ is written as
\begin{align}
E(r_{12})=-g_0^2 \sum_{a} q_1^a q_2^a \biggl[ G(\bm{x}_1|\bm{x}_2) -\frac{1}{3\pi^2} g_0^2 \Lambda^2 \int d^3 z G(\bm{x}_1|\bm{z}) G(\bm{z}|\bm{x}_2)+\mathcal{O}(\Lambda^3) \biggr].
\end{align}
As is stated in the introduction, the expansion  begins with the convolution of the two Coulomb Green functions 
 and is non-singular at the short distance limit of the two external sources $\bm{x}_1=\bm{x}_2$. 
 Up to the same order in the expansion, this term does not appear in QED.

\section{Conclusion}

In this paper, we revisited the non-abelian static force potential in the well-known Coulomb gauge at a finite periodic/twisted box in perturbation theory. Exploiting the finite box as an infrared cutoff, we have given both derivative and short-distance expansions in position space. The former expansion has given us a non-abelian Uehling potential in the next leading order. Also, we have written the interaction energy $E (r_{12})$ \eqref{E12} in the twisted box, explicitly. As a result, the role of the twist on the Green function has been clarified.

\section*{Acknowledgment}
We are grateful to Takuya Hirose, Kazunobu Maruyoshi and Hidenori Sonoda for helpful comments. The work of HI is supprted in part by JSPS KAKENHI Grant Number 19K03828 and by the Osaka City University (OCU) Strategic Research Grant 2020 for priority area. The work of TF is supported by Grant-in Aid for JSPS Fellows $\#$20J15045.

\appendix

\section{Poisson resummation formula}
In this appendix, we give the proof for the Poisson resummation formula in three dimensions:
\begin{align}
	\sum_{\bm{n}\in \mathbb{Z}^3} f(\bm{n})=\sum_{\bm{m}\in \mathbb{Z}^3} \tilde{f}(2\pi\bm{m}),
	\label{Poisson_resum}
\end{align}
where $\tilde{f}$ is the Fourier transform of function $f$:
\begin{align*}
	\tilde{f}(\bm{k})=\int d^3 r f(\bm{r}) e^{-i\bm{k}\cdot\bm{r}}.
\end{align*}
The proof exploits the formula
\begin{align}
	\sum_{\bm{n}\in \mathbb{Z}^3} e^{2\pi i \bm{k}\cdot\bm{n}}
	=\sum_{\bm{m}\in \mathbb{Z}^3} \delta^{(3)}(\bm{k}-\bm{m}),
	\label{delta_formula}
\end{align}
and goes as 
\begin{align*}
	\sum_{\bm{n}\in \mathbb{Z}^3} f(\bm{n})
	&=\sum_{\bm{n}\in \mathbb{Z}^3} \int\frac{d^3 k}{(2\pi)^3} \tilde{f}(\bm{k}) e^{i\bm{k}\cdot\bm{n}}
	=\int\frac{d^3 k}{(2\pi)^3} \tilde{f}(\bm{k}) \sum_{\bm{n}\in\mathbb{Z}^3} e^{i\bm{k}\cdot\bm{n}}\\
	&=\int\frac{d^3 k}{(2\pi)^3} \tilde{f}(\bm{k}) \sum_{\bm{m}\in\mathbb{Z}^3} \delta^{(3)}(\bm{k}-\bm{m})
	=\sum_{\bm{m}\in \mathbb{Z}^3} \tilde{f}(2\pi\bm{m}).
\end{align*}

We can generalize \eqref{Poisson_resum}  by considering 
$\bm{n}\in\mathbb{Z}^3+\bm{\lambda}$ ($\bm{\lambda}\in\mathbb{R}^3$) in the sum in the left hand side:
\begin{align}
	\sum_{\bm{n}\in \mathbb{Z}^3 +\bm{\lambda}} e^{2\pi i \bm{k}\cdot\bm{n}}
	=\sum_{\bm{m}\in \mathbb{Z}^3} \delta^{(3)}(\bm{k}-\bm{m}) e^{2\pi i \bm{\lambda}\cdot\bm{m}}.
\end{align}
This leads us to
\begin{align}
	\sum_{\bm{n}\in \mathbb{Z}^3+\bm{\lambda}} f(\bm{n})
	=\sum_{\bm{m}\in \mathbb{Z}^3} \tilde{f}(2\pi \bm{m}) e^{2\pi i \bm{\lambda}\cdot\bm{m}},
\end{align}
which is exploited in the text.

\end{document}